\documentclass[12pt]{article}
\usepackage[utf8]{inputenc} 
\DeclareUnicodeCharacter{1EF3}{\`y}
\usepackage[letterpaper, margin=1in]{geometry}
\usepackage{graphicx}
\usepackage[style=nature,backend=biber]{biblatex}
\usepackage{hyperref} 
\usepackage{xr-hyper}
\usepackage{float}
\bibliography{references}
\RequirePackage[T1]{fontenc} \RequirePackage[tt=false, type1=true]{libertine} \RequirePackage[varqu]{zi4} \RequirePackage[libertine]{newtxmath}
\usepackage{caption}
\usepackage{subcaption}
\usepackage{setspace}
\usepackage{authblk}
\usepackage{lineno}
\usepackage{booktabs}
\usepackage{multirow}
\usepackage{amsmath}

\RequirePackage[normalem]{ulem}
\RequirePackage{color}
\definecolor{RED}{rgb}{1,0,0}
\definecolor{BLUE}{rgb}{0,0,1}

\title{Social learning drives underprioritization of collective challenges}

\author[a]{Russ Yoon}
\author[b, a]{Vicky Chuqiao Yang}

\affil[a]{Institute for Data, Systems, and Society, Massachusetts Institute of Technology, Cambridge, MA 02139}
\affil[b]{MIT Sloan School of Management, Massachusetts Institute of Technology, Cambridge, MA 02139}

\date{}

\begin{document} 

\maketitle

\begin{abstract}

Societies often struggle to prioritize important challenges in a timely manner, with substantial costs from delayed action on issues like climate change and pandemic mitigation. A persistent puzzle is that broad concern on issues often fails to translate into collective priority. We argue that a key driver lies in how concern is formed across competing issue domains. Some issues depend heavily on social learning, where individuals infer importance from others, often because direct experience is limited. Others depend more on individual learning from firsthand experience. We develop a dynamic model in which two subgroups form issue-specific concerns through individual and social learning, and these concerns are aggregated into collective priority. The model yields three insights. First, with two issues of equal objective severity, the one that depends more on social learning tends to be underprioritized when both issues are severe. Second, gradual increases in severity delay reprioritization of the issue, with the delay growing as reliance on social learning increases. Third, this bias can be reduced by reducing social learning or by increasing intergroup learning beyond a critical threshold. These results offer a general mechanism for why severe problems can remain neglected in collective action despite widespread concern, and why intergroup interaction or experiential simulations may help align collective priorities with objective risks.

\end{abstract}

\section*{Significance statement}

Societies must continually decide which collective challenges to act on, yet they often neglect problems that are widely recognized as serious. Prevailing explanations point to external forces such as media or elite influence. We show that underprioritization can arise from a more basic source: how people form concern in the first place. Using a simple mathematical model, we find that when issues make competing demands, those whose importance is learned mainly socially---from others rather than from direct experience---can be systematically underprioritized even when equally severe. Rising severity may correct this only slowly. The result offers a general account for why some threats mobilize action while others languish, along with the conditions under which that gap can close. 

\section{Introduction}

Human societies often face multiple, sometimes conflicting, challenges simultaneously. We seek economic growth while striving to mitigate climate change, or aim to curb the spread of an epidemic through social distancing while attempting to protect retail businesses. Yet efforts to balance competing priorities frequently fall short. Despite scientific consensus on the threat of anthropogenic climate change \cite{lee2023ipcc, cook2016consensus} and a majority of Americans expressing concern \cite{leiserowitz2025climate}, no consistent federal-level mitigation policy has emerged. Similarly, during the COVID-19 pandemic, partisan divisions undermined the effectiveness of containment measures such as social distancing. A recurring pattern in these failures is that collective decision-making often falters, failing to prioritize societal threats in a timely manner.

\begin{figure}[H]
    \centering
    \includegraphics[width=\linewidth]{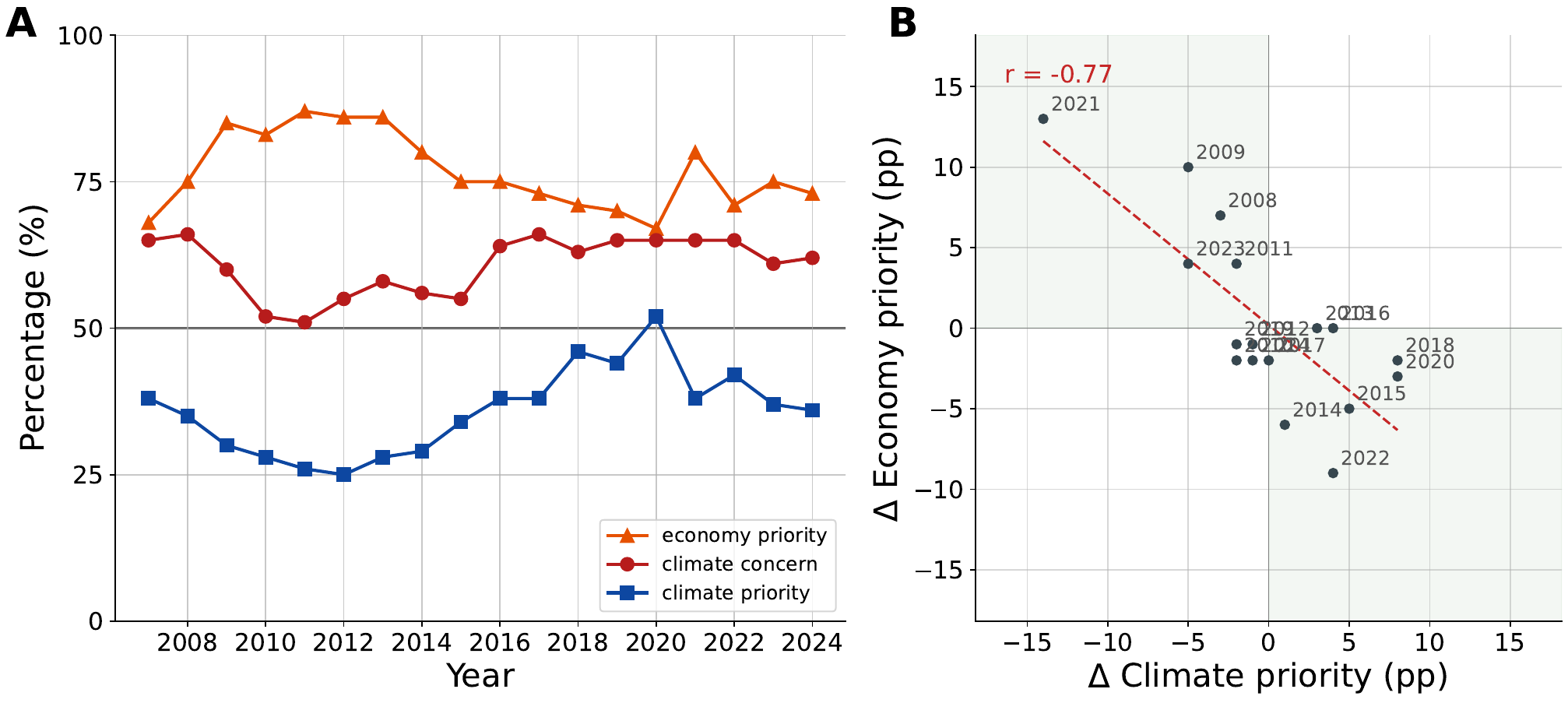}
    \caption{Majority concern over an issue does not translate into majority priority. (A) Percentage of adults expressing concern about climate change (Gallup) and percentage stating climate change or the economy should be a top national priority (Pew), among U.S. adults, 2007--2024. (B) Year-over-year changes in climate and economy priority ($r = -0.77$).}
    \label{fig:survey}
\end{figure}

Figure~\ref{fig:survey} illustrates this pattern in U.S. public opinion. The survey data consistently show that roughly 60\% of Americans report worrying about global warming, yet only about 35\% rank climate change mitigation as a top national priority, far below the economy at roughly 75\% (Fig.~\ref{fig:survey}A). A majority of Americans are concerned about climate change, but a minority treat it as a top governmental priority. Because climate mitigation and economic growth involve inherent trade-offs, prioritizing one necessarily comes at the cost of the other. For example, policies that curb emissions tend to slow near-term economic expansion, while growth-oriented policies typically increase emissions. Indeed, the data show that year-over-year changes in climate and economy priority are strongly negatively correlated ($r = -0.77$, Fig.~\ref{fig:survey}B). The persistent failure to prioritize climate action has been widely attributed to exogenous forces such as partisan elite cues and strategic media framing \cite{merkley2018party,bohr2020reporting}. Here, we propose a complementary, endogenous explanation grounded in the psychological processes of concern formation, and we use this framework to evaluate interventions.

The formation of opinions in collectives has been widely studied using dynamical models across physics, mathematics, psychology, and sociology \cite{castellano2009statistical,hickok2022bounded,dalege2025networks}. Most models analyze how opinions about a single issue evolve under social influence and external signals, with some extending to multi-dimensional opinion vectors \cite{schweighofer2020agent,baumann2021emergence,parsegov2016novel}. A smaller but growing literature examines cross-issue interactions, emphasizing how associations between issues are reinforced through social interaction \cite{dellaposta2015liberals,goldberg2018beyond}. Yet it remains unclear how concerns for multiple issues with distinct characteristics compete for priority in a polarized society, a problem central to the emerging discourse on collective adaptation \cite{galesic2023beyond}.

Opinion formation is typically attributed to two primary mechanisms---social learning and individual learning \cite{boyd1988culture}. Social learning refers to adopting the beliefs or behaviors of other people, such as by observing their behaviors or communicated concerns. Here we focus on frequency-based social learning, whereby individuals disproportionately adopt the concerns expressed by the majority of their neighbors. Conformity to majority opinion is a robust and widespread phenomenon in human judgment \cite{capuano2024systematic}, and perceived social consensus is a well-documented driver of concern formation, particularly for issues like climate change \cite{lewandowsky2019science}. By contrast, individual (or asocial) learning occurs when people interact directly with their environment and acquire information that informs their concern about an issue's severity. 

Individuals may use a combination of individual and social learning \cite{degroot1974reaching, galesic2021integrating}, with contextual factors shaping the extent to which each is relied upon. Individual learning is most feasible when the environment is observable, whereas limited observability can push people toward greater reliance on social learning \cite{ganuthula2024limits, kendal2018social}. Other empirical evidence shows that reliance on social learning increases when tasks are difficult or complex \cite{gino2007effects,acerbi2016social,toyokawa2019social,morgan2012evolutionary}, when uncertainty is high \cite{ciranka2020bayesian,toelch2014individual,muthukrishna2016and}, when confidence in one's own judgment is low \cite{pescetelli2021confidence,morgan2012evolutionary}, and when the cost of individual exploration is high \cite{morgan2012evolutionary}. Consequently, the degree of social learning should vary in concern formation across issues, reflecting issue-specific levels of complexity, uncertainty, confidence, and learning costs. We therefore model reliance on social learning as issue-specific.

Climate change and early-stage pandemics exemplify domains where social learning dominates concern formation. Climate impacts are temporally delayed and geographically distant, limiting individual learning and elevating uncertainty \cite{weber2006experience}, and acquiring scientific information about severity is costly for most individuals. Accordingly, climate concern in the United States is strongly structured by political affiliation, with partisanship and ideology frequently outweighing experiential exposure \cite{binelli2023explaining, egan2017climate}. The early COVID-19 pandemic exhibited similar conditions. The virus's invisibility, delayed feedback, and rapidly evolving science made individual learning difficult and unreliable, and concern and protective behavior were strongly shaped by ideology \cite{allcott2020polarization,gollwitzer2020partisan}. Individuals with greater direct adverse experience of COVID-19 exhibited a narrower partisan gap in concern and behavior \cite{constantino2022personal}. Underlying these barriers is that climate concern does not follow from simply perceiving temperature, but from grasping a causal chain linking human activity to warming and its distributed harms.

By contrast, economic concerns---prices, wages, and employment---are directly observable. Individuals can register worry about rising prices or lost income without understanding their macroeconomic causes, reducing the cost of individual learning. As a result, economic concerns are more likely to be shaped by individual learning and less by social learning. Consistent with this view, inflation beliefs co-move strongly with household income changes and direct price experience \cite{taubinsky2024beliefs,cavallo2017inflation}. Although partisan bias has some effects on evaluations of the national economy, its effect on assessments of household finances is marginal \cite{ang2022partisanship,healy2017digging}.

Beyond the degree to which individuals rely on social learning, whom they learn from is also important. In a polarized society, people tend to learn disproportionately from ingroup partisans. This tendency may arise both from social environments that are politically homogeneous \cite{barbera2015tweeting,conover2012partisan} and from affective polarization, which reduces trust in and receptiveness to information provided by outgroup partisans \cite{iyengar2019origins}. Efforts to reduce affective polarization often aim to promote meaningful cross-party collaboration \cite{woodley2025defusing}, encouraging consideration of out-partisan perspectives.

Here, we develop a two-group, two-issue dynamical model of collective issue prioritization that captures competition between issues whose concern formation relies on social learning to differing degrees. Our model yields three main findings. First, when two equally severe issues make competing demands, the issue whose concern formation relies more heavily on social learning is systematically underprioritized. Second, when objective severity increases gradually, collective reprioritization is delayed, and the delay lengthens with reliance on social learning. Third, this underprioritization can be corrected by reducing reliance on social learning or by restoring intergroup learning, though the latter works only above a threshold---so modest depolarization efforts may be insufficient once concern is deeply entrenched.

\section{The Mathematical Model}

We consider a collective composed of two equal-sized subgroups, 1 and 2 (e.g., political parties or other salient identities), facing two competing issues, \textit{H} and \textit{L}. Motivated by examples such as climate change and economic conditions discussed above, issue \textit{H} is one for which concern formation relies more heavily on social learning, whereas issue \textit{L} relies more heavily on individual learning. The two issues are in conflict in the sense that actions advancing one tend to inhibit the other. 

The model has two components. First, individuals form concern about each issue through a combination of individual and social learning. This reflects the fact that political and social issues often develop their own discourse: people encounter issue-specific information, arguments, and norms through personal experience and interaction with others. Social learning can occur both within and across groups, with the extent of intergroup learning governed by a model parameter. Second, collective priority is determined by comparing the total concern devoted to the two competing issues across the whole population. This captures settings in which collective decisions are made through representative or direct democratic processes, where outcomes depend not only on concern within any one subgroup but on the aggregate level of concern across subgroups.

We assume that each issue has an objective severity, denoted as $I_i$, which is between 0 and 1. We consider this objective severity informs individual learning. Under purely individual learning, the expected proportion of individuals concerned about issue $i$ should increase with severity $I_i$. For simplicity, we assume the dependence is linear---the proportion concerned about issue $i$ is $I_i$---and is the same across the two groups.

We begin by specifying the level of concern that individuals would adopt given the information available to them. Individuals can learn about an issue in two ways: individual learning or social learning. Concern formation is modeled as a weighted combination of these two sources of information.

The parameter \(s_i\) governs the relative weight placed on social versus individual learning for issue \(i\). Taking the expectation over individuals yields the target level of concern within a group:
\begin{equation}\label{eq:target}
C^*_{g,i}(t)=(1-s_i)\,I_i+s_i\,f\!\big(X_{g,i}(t);\alpha\big).
\end{equation}

\begin{figure}[t!]
  \centering
\includegraphics[width=\linewidth]{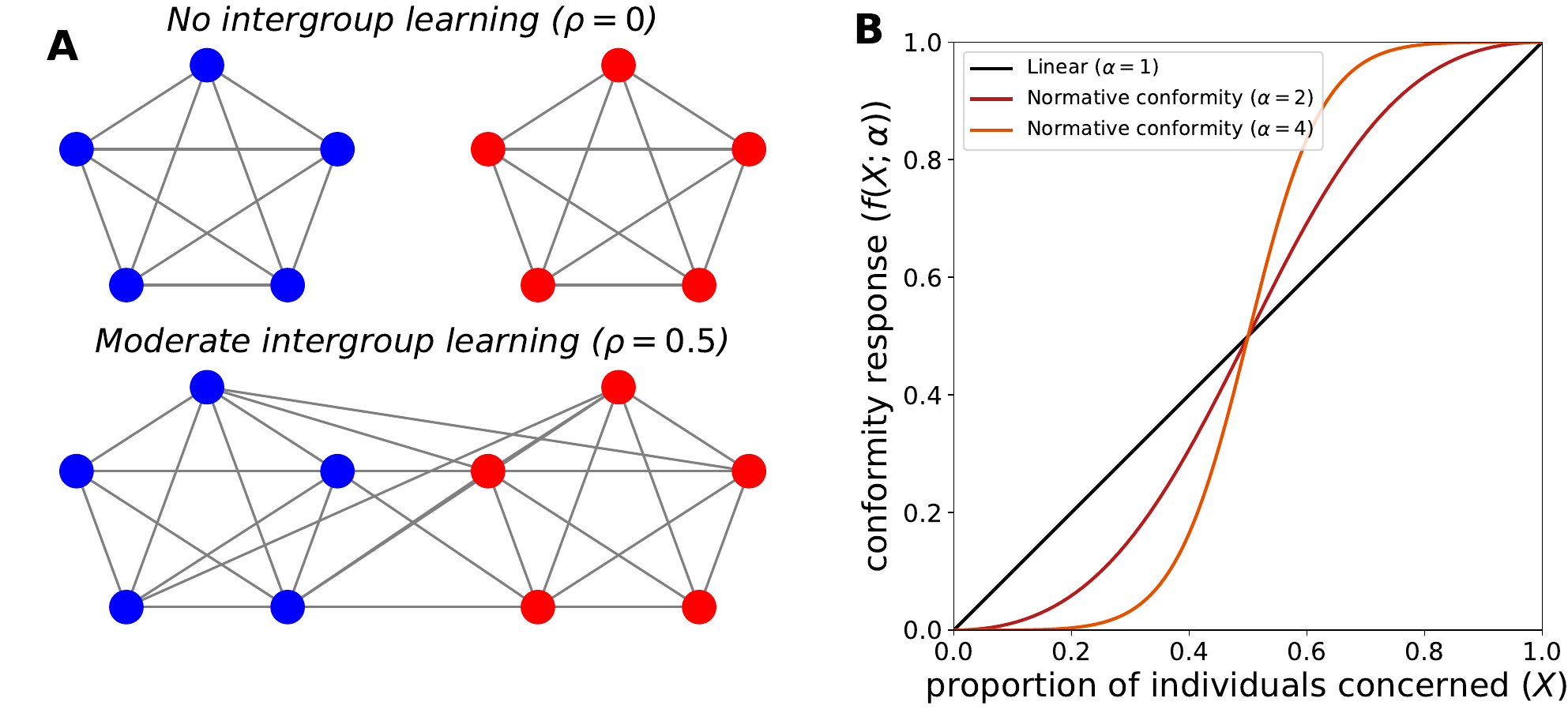}
\caption{Model components. (A) Two-group network with full within-group connectivity (learning). Top panel shows no intergroup learning ($\rho=0$) and bottom panel shows moderate intergroup learning ($\rho=0.5$). Blue and red nodes represent the two groups. (B) Conformity response function $f(X;\alpha)$ for varying levels of normative conformity.}
  \label{fig:model}
\end{figure}

Concern reflects a weighted combination of what individuals learn from their own experience and what they infer from the concerns expressed by others. When \(s_i=0\), concern is determined entirely by objective severity. When \(s_i=1\), concern is determined entirely by the observation of others. The first term represents concern generated through individual learning, where \(I_i\) denotes the objective severity of issue \(i\). The second term represents concern generated through social learning. We focus on frequency-dependent social learning, an important pathway through which individuals use the prevalence of a belief or behavior among others as information \cite{henrich1998evolution}. In this form of social learning, majority signals are often amplified: the likelihood of adoption often exceeds its prevalence level in the population for majority belief or behavior \cite{claidiere2012integrating,claidiere2014frequency}. We capture this process by transforming the observed prevalence of concern among peers by group $g$, \(X_{g,i}(t)\), through an S-shaped conformity response function, $f\!\big(X_{g,i}(t);\alpha\big)$, with shape parameter $\alpha$ (Fig.~\ref{fig:model}B). 

Group-level concern adjusts gradually rather than instantly. It moves toward the target $C^*_{g,i}$ at a rate set by the timescale $\tau$, reflecting the inertia \cite{sterman2002system},
\begin{equation}\label{eq:dynam}
\frac{d C_{g,i}(t)}{dt}=\frac{C^*_{g,i}(t)-C_{g,i}(t)}{\tau}.
\end{equation}

The observed concern \(X_{g,i}(t)\) can come from either ingroup or outgroup members. We capture the degree of intergroup learning with a single parameter, $\rho$, which is the probability that an individual incorporates concern from a given outgroup member, with the probability of learning from ingroup members normalized to 1. This abstraction is not intended to reproduce the full structure of real social networks, but to isolate how affective polarization affects socially learned concerns. When $\rho=0$, individuals learn only from ingroup members, representing a highly polarized society (Fig.~\ref{fig:model}A, upper panel). When $\rho=1$, outgroup members affect one as much as ingroup members. The observed concern, \(X_{g,i}(t)\), is therefore a weighted average of the two groups. For group 1, it is
\begin{equation}\label{eq:X}
X_{1,i}(t)=\frac{1}{1+\rho}\,C_{1,i}(t)+\frac{\rho}{1+\rho}\,C_{2,i}(t),
\end{equation}
with the analogous expression for group 2. This completes the derivation for the dynamics of concern formation.

Given group-level concerns, assuming the two groups are equal in size, population-level concern for issue \(i\) is the average of the two group concerns, \( C_{i}(t)=\tfrac{1}{2}C_{1,i}(t)+\tfrac{1}{2}C_{2,i}(t) \). The collective prioritizes \textit{issue H} relative to \textit{issue L} by comparing these two levels of concern. We use a logit choice formulation, a standard representation of probabilistic choice between competing alternatives, to map relative concern into collective priority \cite{ziegler2012individual,vsvcasny2017public,berger2022willingness,streimikiene2019review,sterman2002system, mcfadden1974conditional}. Under this rule, the issue with greater concern is more likely to be prioritized, and the strength of this tendency increases with the concern gap. The probability that issue $H$ is prioritized compared to $L$ is,
\begin{equation}\label{eqn:logit}
P(t)=\frac{\exp\!\big(C_{H}(t)/\sigma\big)}{\exp\!\big(C_{H}(t)/\sigma\big)+\exp\!\big(C_{L}(t)/\sigma\big)},
\end{equation}

where \(\sigma>0\) sets how sharply the collective responds to a gap in concern. When \(\sigma\) is small, even a slight advantage in concern translates into near-complete priority. When \(\sigma\) is large, priority responds only gradually.

\section{Results}

When two equally severe issues compete for collective attention, we show the issue whose concern formation relies more heavily on social learning is systematically underprioritized at equilibrium. To illustrate this mechanism, we begin with two fully disconnected groups, representing an extreme case of affective polarization. We start with a simple case where two issues $H$ and $L$ have the same objective severity ($I$), while concern formation about $H$ depends more strongly on social learning.

\begin{figure}[H]
    \centering
    \includegraphics[width=0.6\linewidth]{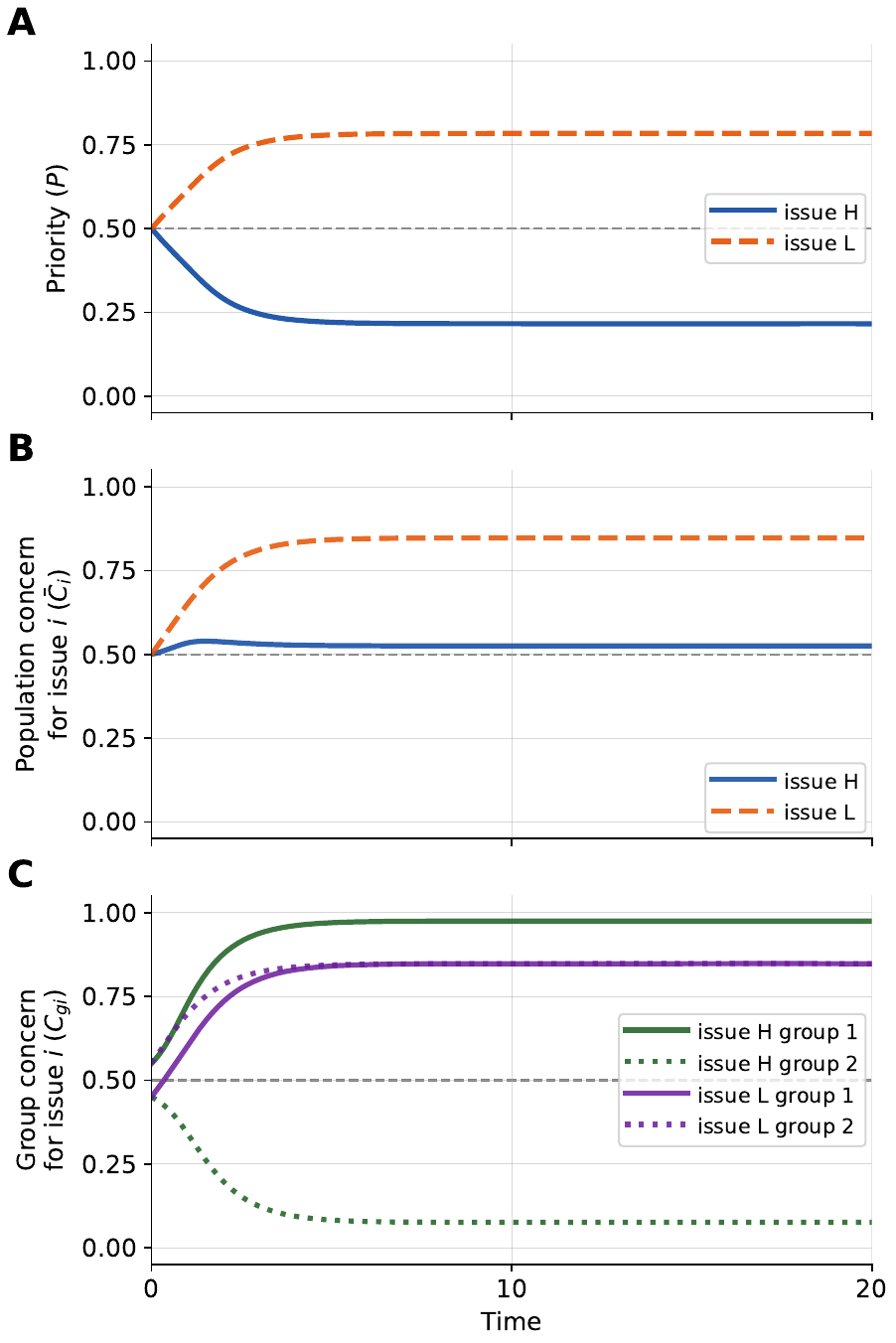}
      \caption{High social learning produces concern polarization and underprioritization. (A) Population priority for $H$ and $L$. (B) Population-level concern for each issue. (C) Group-level concern for $H$, showing between-group divergence.}
    \label{fig:dynamics}
\end{figure}

The model predicts that issue $L$ reaches higher priority than issue $H$ (Fig.~\ref{fig:dynamics}A), despite having the same objective severity. This outcome occurs because concern for $H$ is concentrated within a single group, while the other remains largely unconcerned (Fig.~\ref{fig:dynamics}C). In contrast, concern for $L$ is shared across both groups, albeit at an intermediate level. In this particular simulation, a majority of the population becomes concerned about $H$ (Fig.~\ref{fig:dynamics}B). Thus, a majority of individuals being concerned about an issue does not necessarily translate into prioritizing that issue due to trade-offs with another issue. In this trade-off, issues that rely more heavily on social learning can be disadvantaged in collective prioritization because social dynamics concentrate concern within groups rather than distributing it across the population. We call this underprioritization due to social learning, \textit{collective prioritization bias}. 

This outcome provides a potential explanation for empirical cases in which widespread concern fails to translate into collective priority. For example, most Americans express concern about climate change, and scientific assessments indicate that it poses a severe societal risk. Yet climate mitigation often receives lower priority than economic issues. The model reproduces this qualitative pattern: a socially learned concern can command majority support overall while remaining concentrated within political groups. A competing issue that is grounded more in immediate experience, and therefore more evenly distributed across groups, can then dominate collective priorities.

\begin{figure}[h!]
  \centering
\includegraphics[width=\linewidth]{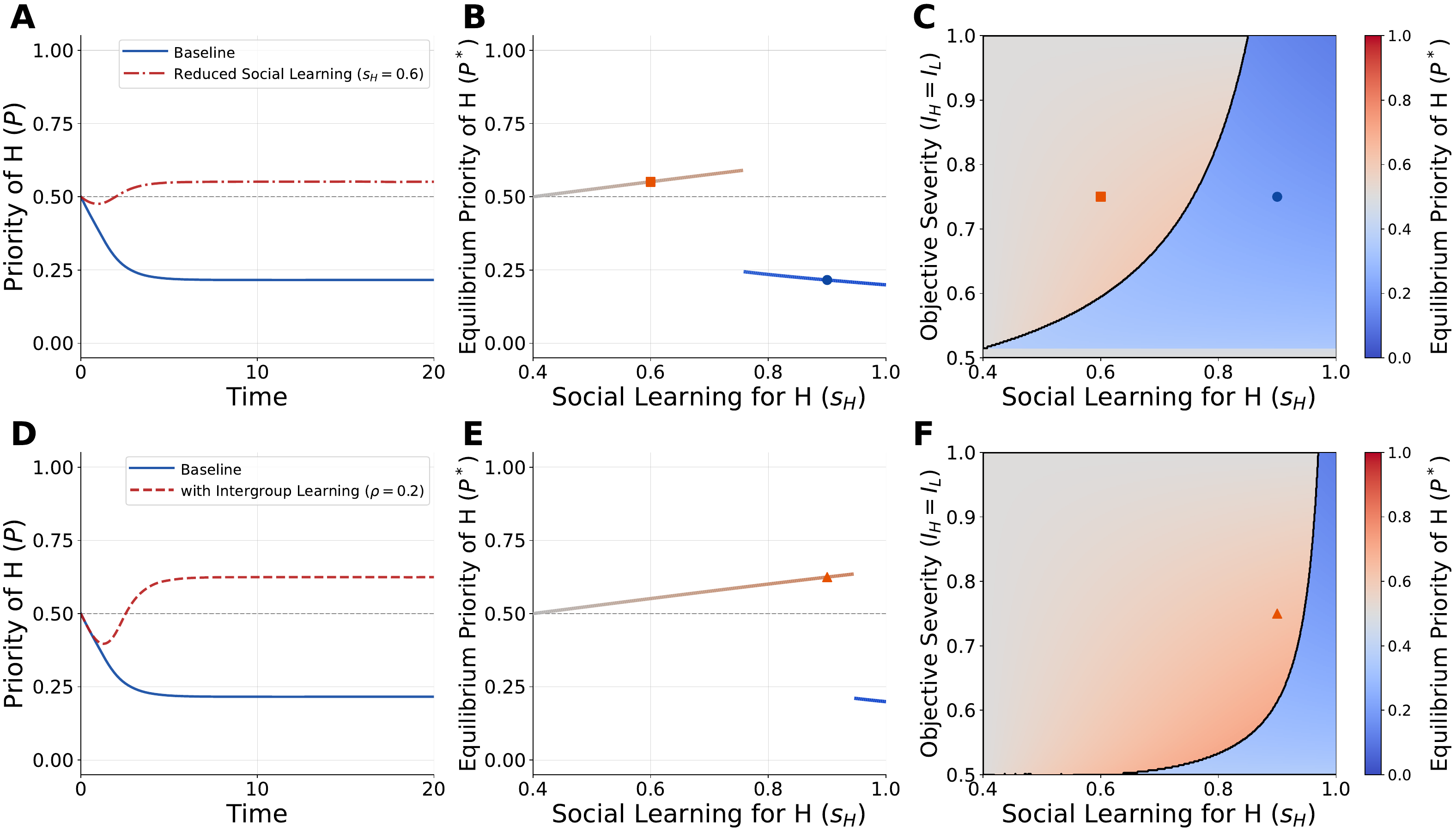}
  \caption{Both reduced social learning and intergroup learning can restore prioritization of $H$. (A) Reducing social learning for $H$ ($s_H=0.6$) raises its priority relative to the baseline. (B) Equilibrium priority of $H$ across $s_H$ under the conditions in A. (C) Equilibrium priority across $s_H$ and shared objective severity ($I_H=I_L$) without intergroup learning. (D) Introducing intergroup learning ($\rho=0.2$) raises the priority of $H$ relative to the baseline. (E) Equilibrium priority of $H$ across $s_H$ with intergroup learning. (F) Equilibrium priority across $s_H$ and shared objective severity with intergroup learning. Markers indicate the equilibria in A (square, circle) and D (triangle).}
  \label{fig:phase}
\end{figure}

The preceding result suggests that underprioritization of $H$ is not driven by its objective severity, but by the way concern for it is formed and distributed across groups. Two model parameters directly govern this process. First, reducing social learning for issue $H$ can increase its priority, even to above a majority (Fig.~\ref{fig:phase}A). Second, introducing intergroup learning reduces divergence in concern for $H$ across groups and similarly restores a majority priority for $H$ (Fig.~\ref{fig:phase}D).

The collective prioritization bias emerges only once social learning is strong enough. Reducing social learning for $H$ raises its priority relative to the baseline (Fig.~\ref{fig:phase}A). Generalizing this single comparison, we compute the equilibrium priority of $H$ across levels of social learning ($s_H$) at fixed objective severity (Fig.~\ref{fig:phase}B). Priority does not vary smoothly with $s_H$. For weak social learning it rises slightly above parity, as social reinforcement modestly amplifies shared concern, but once $s_H$ crosses a critical threshold it drops abruptly and $H$ becomes a minority priority. The circle and square mark the baseline (simulation in Fig.~\ref{fig:dynamics}) and reduced-$s_H$ equilibria from panel A, locating the minority-to-majority transition at this threshold. Generalizing further across shared objective severity ($I_H = I_L$) yields the heat map in Fig.~\ref{fig:phase}C. The blue region marks the conditions under which $H$ is underprioritized relative to $L$, showing that the threshold, and the bias beyond it, persist across the full severity range.

Intergroup learning counteracts the bias by shifting the same threshold, and it does so across the full severity range. Introducing it raises the priority of $H$ relative to the baseline (Fig.~\ref{fig:phase}D) by raising the critical threshold in $s_H$ (Fig.~\ref{fig:phase}E), which shrinks the underprioritization region across the severity range (Fig.~\ref{fig:phase}F). The parameter point that lies in the minority region without intergroup learning (circle, panel C) falls in the majority region once it is introduced (triangle, panel F). Intergroup learning therefore improves collective prioritization across a wide range of severities, not only at the single point examined in the dynamics.

Panels C and F restrict the objective severity axis to the upper half of the parameter range, where a majority of the population would be concerned under purely individual learning. This regime captures competition among severe societal challenges. At lower objective severity, the pattern reverses: social learning leads $H$ to be overprioritized. The full heat map over the complete severity range is shown in the Supplementary Information.

\begin{figure}[h]
  \centering
  \includegraphics[width=.8\linewidth]{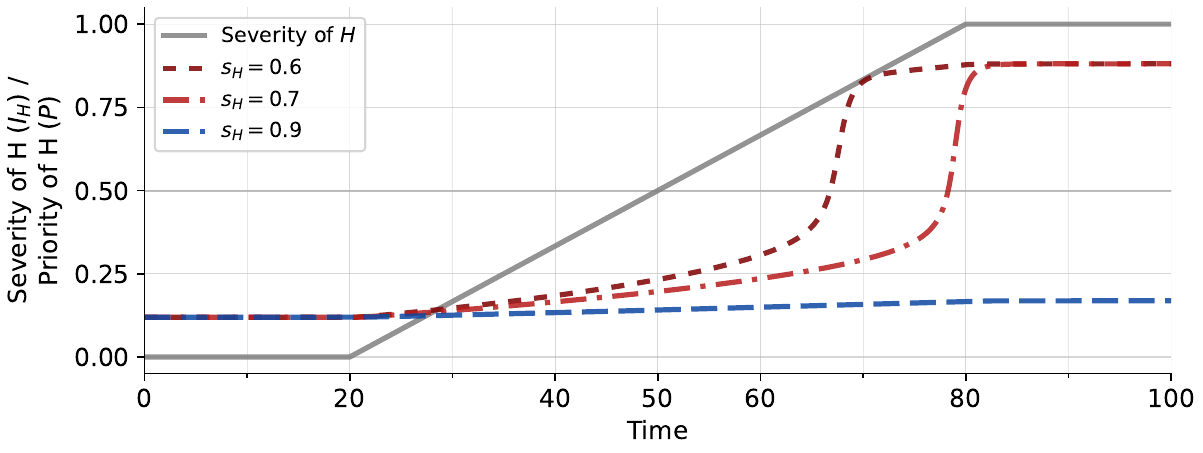}
    \caption{Higher social learning delays prioritization of issue $H$. The gray line shows the objective severity of $H$, rising linearly until reaching $1$, while $L$ is held at $0.5$. Colored lines show the priority of $H$ over $L$ at three levels of social learning. Stronger social learning lengthens the delay before priority reaches equilibrium.}   \label{fig:delay}
\end{figure}

The phase diagrams characterize equilibrium priorities at fixed severity, but many societal risks intensify over time. We therefore ask whether collective priorities adjust as objective severity increases gradually. Fig.~\ref{fig:delay} shows trajectories in which the objective severity of $H$ rises over time while that of $L$ remains fixed, for several levels of social learning. Social learning alone sets the equilibrium priority given the ultimate severity: it is the same whether severity rises gradually from a low value or starts at its maximum (Supplementary Information). The gradual increase in severity affects only the timing, delaying when priority reaches that equilibrium. Stronger reliance on social learning lengthens this delay.

When social learning is high and severity is reached gradually, early low-concern equilibria persist through social reinforcement, leaving the issue underprioritized well after the severity of $H$ surpasses that of $L$. Issues that worsen gradually and depend more on social learning are therefore especially prone to delayed prioritization.

\begin{figure}[h]
  \centering
  \includegraphics[width=.8\linewidth]{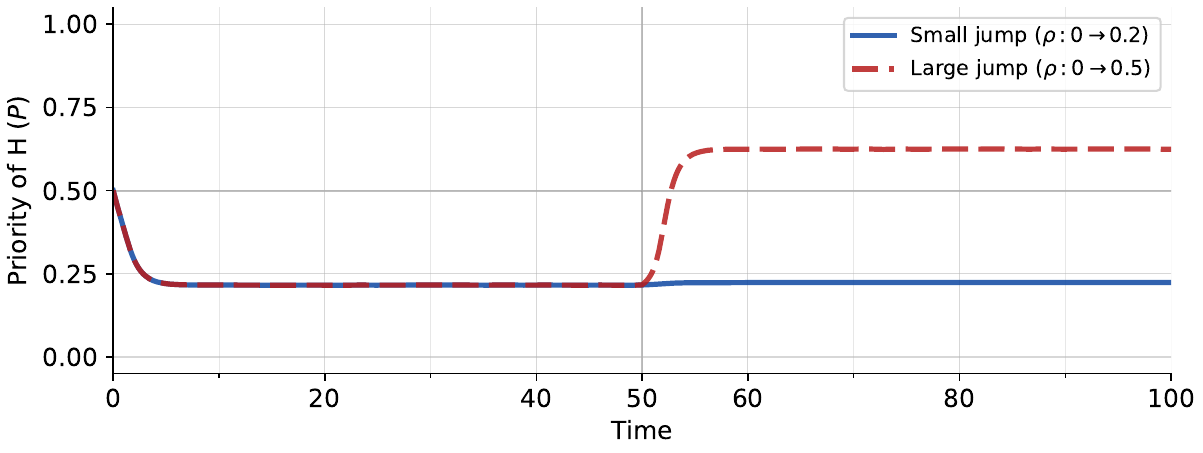}
    \caption{Restoring intergroup learning corrects underprioritization only above a threshold. The two scenarios differ in the size of the increase in intergroup learning at $t = 50$.}
  \label{fig:dynamic_connectivity}
\end{figure}

We next ask whether increasing intergroup learning, such as through cross-party dialogue initiatives, can be an effective intervention against this bias after concern has already polarized. Figure~\ref{fig:dynamic_connectivity} compares two scenarios that begin from the baseline simulation in Fig.~\ref{fig:dynamics}, run until the system settles into an equilibrium with polarized concern and minority priority for \textit{H}, and then raise intergroup learning at $t=50$ to different degrees. A small increase in intergroup learning leaves these outcomes essentially unchanged. A larger increase leads the two groups to converge on high concern for $H$, shifting $H$ to majority priority (In Supplementary Information we show the underlying group-level concern dynamics for both issues). Thus, interventions aiming to increase intergroup learning must exceed a threshold to change the collective priority of issues. Small increases in intergroup learning may be insufficient once concern is entrenched.

\section{Discussion}

Our results show how issue characteristics and social learning jointly shape collective prioritization between competing issues. When issues are equally severe, social dynamics can lead groups to underprioritize the one that relies more heavily on social learning. When objective severity increases gradually, collective prioritization of a high--social-learning issue can shift only after a substantial delay. When intergroup learning increases from a polarized baseline, it must exceed a threshold before concerns converge and priorities realign with severity. Below that threshold, intergroup interaction produces only modest reductions in polarization. Notably, this bias is distinct from individual-level cognitive biases \cite{weber2006experience}. Our model assumes that individuals, when using individual learning, on average accurately perceive objective severity, yet collective underprioritization emerges from social dynamics alone.

The model's predictions align with the puzzling patterns in U.S. public opinion. Climate change mitigation has been de-prioritized behind economic policies in spite of the scientific consensus on the severity of anthropogenic climate change and a majority of Americans reporting being concerned about climate change. The gap between majority concern for climate change and its minority standing as a national priority, relative to the economy (Fig.~\ref{fig:survey}A), is what the model produces when a high--social-learning issue competes with a low--social-learning issue of comparable objective severity. 

The present model is built on inherent policy trade-offs, where actions taken to mitigate one issue directly exacerbate the other, as when emissions reduction raises energy costs and slows near-term economic growth. However, the underlying logic may generalize to issues that are not inherently in conflict, because individuals also face a finite cognitive budget for attention \cite{evensen2021effect,sisco2023examining}. To the extent that attention allocated to one issue reduces attention available for others, similar underprioritization dynamics could emerge even between issues whose objectives do not directly oppose each other.

The model also predicts that collective responses to different issues can unfold at markedly different speeds. When objective severity increases gradually for a high--social-learning issue, concern and priority may adjust only after a substantial delay. The model suggests that such variation can arise in part from differences in how much concern formation depends on social learning relative to direct experience. Importantly, the responsiveness of collective adjustment may also vary across communities, reflecting differences in social network structure and cultural tightness \cite{gavrilets2026cultural} that shape how strongly individuals rely on ingroup cues.

A further implication concerns politicization. When a high--social-learning issue becomes tied to group identity and intergroup learning is near zero, within-party advocacy alone cannot shift collective prioritization. If the science-communication environment turns climate beliefs into markers of cultural identity \cite{kahan2015climate}, advocacy within one party can reinforce rather than reduce the concern gap. Restoring intergroup learning is a prerequisite for collective reprioritization, unless both parties independently arrive at concern through separate channels. This contrasts with the ozone-depletion episode, which was resolved amid lower politicization, greater immediate salience, aligned elite messaging, and the availability of commercially viable chlorofluorocarbon (CFC) substitutes \cite{sunstein2007montreal, grundmann2006ozone, baldwin2020solving}.

\begin{figure}[t!]
    \centering
    \includegraphics[width=.8\linewidth]{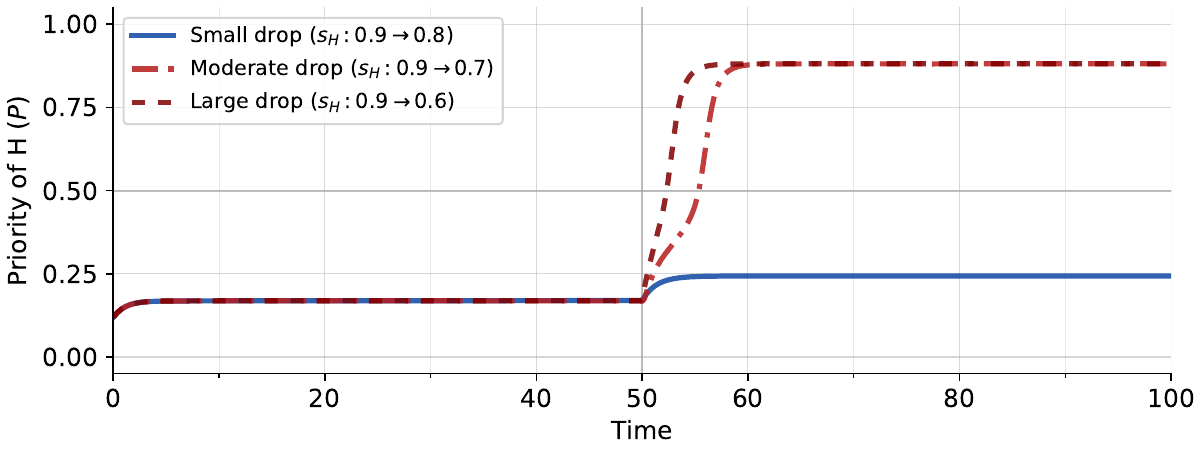}
    \caption{Reducing social learning can shift collective priority toward the high--social-learning issue. Starting from the high--social-learning equilibrium in Fig.~\ref{fig:delay}, $s_H$ is reduced at $t=50$. Larger reductions produce more responsive shifts in priority.}
    \label{fig:s_drop}
\end{figure}

The model highlights two levers for shifting collective priorities. The first is reducing the degree to which concern formation depends on social learning. Figure~\ref{fig:delay} showed that with sufficiently high social learning, priority settles at a low equilibrium even after severity reaches its maximum, so raising severity alone cannot reprioritize the issue. Starting from that outcome, Fig.~\ref{fig:s_drop} shows that reducing $s_H$ can unlock collective reprioritization, with larger reductions producing faster and more complete shifts. This suggests that interventions making the causal chain linking human activity to changes in temperature and environment more experiential and concrete, such as interactive climate simulators \cite{rooney2021building}, can complement cross-partisan engagement by shifting the basis of concern from social learning toward individual learning. These two levers operate through distinct channels, and combining both may be more effective than either alone.

The second lever is increasing intergroup learning. When baseline intergroup learning is very low, small increases have little effect, but sustained increases can eventually cross a tipping point, producing a sudden and large shift in concern and collective priorities. This has implications for cross-partisan engagement programs \cite{woodley2025defusing,landry2023intergroup}, which may show limited measurable impact initially but could trigger rapid change once a critical level of intergroup interaction is reached.

The model is intentionally parsimonious to isolate the effects of the key social dynamics. It does not model awareness diffusion, which future work could integrate with the concern-formation dynamics studied here. The framework best fits contexts in which two issues are inherently in conflict in the near term, such as climate mitigation versus economic growth. Over longer horizons these trade-offs can become synergistic, as when the Inflation Reduction Act of 2022 bundled climate investment with job creation and energy cost reduction. We model two groups to capture two-party concern polarization in the United States, though multi-party systems in which coalition formation yields a governing bloc and an opposition can be mapped into the same structure. We model two issues because clear direct trade-offs among three or more issues are rarer in practice. Finally, the model assumes frequency-based social learning with homogeneous influence within groups. Incorporating expert learning, heterogeneous credibility, and real-world network structure is an important direction for future work.

Our work makes two contributions. First, we offer an endogenous explanation for patterns widely observed in polarized societies, such as the persistent gap between climate concern and climate prioritization, without attributing these outcomes to media distortion or elite manipulation. The underprioritization instead emerges from the structure of social influence, operating through ingroup and outgroup dynamics. Second, we extend the opinion dynamics literature from single-issue belief formation to multi-issue priority-setting under competing demands. Most formal models analyze how opinions about a single issue evolve under social influence \cite{shin2025culture, castellano2009statistical}, but real-world challenges rarely arise in isolation and often compete. Recent work has called for formal models that address how societies manage multiple competing challenges, notably the collective adaptation framework \cite{galesic2023beyond}. Our model responds to this call by making it possible to study when societies adapt to changing severity across domains and when they underprioritize despite rising objective severity.

\subsection*{Acknowledgements}
We thank Mirta Galesic for helpful discussion. V.C.Y. and R.Y. were partially supported by NSF award 2329988.

\printbibliography

\end{document}


\maketitle

\tableofcontents

\section{Data sources}

Figure~\ref{fig:survey} draws on two U.S. public opinion survey series. Climate concern data are from the Gallup Poll Social Series: Environment, an annual March survey of approximately $1{,}000$ U.S. adults. We use the combined share reporting that they worry ``a great deal'' or ``a fair amount'' about global warming or climate change. Policy priority data are from Pew Research Center, which asks whether each   of a set of issues should be a ``top priority'' for the President and Congress. We use the share rating ``dealing with global climate change'' and ``strengthening the nation's economy'' as top priorities. The series was administered by phone through 2020 ($N \approx 1{,}500$) and via the American Trends Panel online from 2021 onward ($N \geq 2{,}500$ per item). The mode change may introduce a level shift. Toplines are available from Gallup (\url{https://news.gallup.com/poll/1615/environment.aspx}) and Pew Research Center (\url{https://www.pewresearch.org/topic/politics-policy/}).

Panel A of Fig.~\ref{fig:survey} plots all three series over the 18-year window (2007--2024) on which they jointly overlap. Panel B examines their year-over-year comovement. Let $\Delta x_t = x_t - x_{t-1}$ denote the annual first difference of series $x$. Applying this to the Pew climate-priority and economy-priority series over the overlap window yields $n = 17$ paired observations (2008--2024). The Pearson correlation between the two first-difference series is $r = -0.77$ ($p < 0.001$). The dashed red line in Panel B is the ordinary least-squares fit of $\Delta$ Economy priority on $\Delta$ Climate priority ($R^2 = 0.60$), with slope $-0.82$ (SE $= 0.17$, $p < 0.001$) and intercept $0.20$ percentage points (SE $= 0.91$, $p = 0.83$).

\section{Details of model simulation}

Table~\ref{tab:baseline_params} lists the baseline parameter values, and Table~\ref{tab:figure-params} records the parameters used in each figure. These values are chosen to illustrate the model's qualitative regimes and are not calibrated to data. The baseline places the two issues in the regime relevant to the climate--economy case in Fig.~\ref{fig:survey}, with both issues at high objective severity and issue $H$ relying more heavily on social learning. The equilibrium maps in Fig.~\ref{fig:phase}C and F, together with Fig.~\ref{fig:phase_SI}, show that these regimes persist across a broad region of parameter space.

The social learning weights $s_H = 0.9$ and $s_L = 0.4$ instantiate the model's central contrast between an issue whose concern formation relies more on social learning and one that relies more on individual learning. Initial group concerns are small symmetric perturbations around 0.5, allowing within-group reinforcement to amplify slight asymmetries. The direction of the perturbation is reversed across the two groups for the two issues, modeling groups with slightly different initial concerns. The conformity exponent $\alpha = 3$ yields an S-shaped response used to model frequency-based conformity \cite{henrich1998evolution,claidiere2014frequency}, and the choice sensitivity $\sigma = 0.25$ makes prioritization responsive to concern differences without being deterministic. Objective severities $I_H = I_L = 0.75$ place both issues in the high-severity regime relevant to the climate--economy case (Fig.~\ref{fig:survey}A). Setting intergroup learning to $\rho = 0$ represents a maximally polarized baseline, and $t_{\text{final}} = 20$ is well past equilibrium for baseline simulations.

\begin{table}[H]
\centering
\caption{Baseline parameter values used in Fig.~\ref{fig:dynamics}. All other figures inherit these values except where Table~\ref{tab:figure-params} specifies otherwise.}
\label{tab:baseline_params}
\renewcommand{\arraystretch}{1.2}
\begin{tabular}{l c l}
\toprule
Parameter & Value & Description \\
\midrule
$\mathbf{C}_0$              & $(0.55,\,0.45,\,0.45,\,0.55)$ & Initial group concerns, $(C_{1H},\,C_{2H},\,C_{1L},\,C_{2L})$ \\
$s_H,\,s_L$                 & $0.9,\,0.4$ & Social learning weight (issues $H$, $L$) \\
$I_H,\,I_L$                 & $0.75,\,0.75$ & Objective severity (issues $H$, $L$) \\
$\rho$                      & $0$    & Intergroup learning \\
$\sigma$                    & $0.25$ & Logit choice sensitivity \\
$\alpha$                    & $3$    & Conformity exponent \\
$t_{\text{final}}$          & $20$   & Integration horizon \\
\bottomrule
\end{tabular}
\end{table}

\begin{table}[H]
\centering
\caption{Parameter values for each figure that differ from the baseline in Table~\ref{tab:baseline_params}. Parameters not listed are at baseline.}
\label{tab:figure-params}
\footnotesize
\renewcommand{\arraystretch}{1.25}
\begin{tabular}{l p{0.72\linewidth}}
\toprule
Figure & Changes from baseline \\
\midrule
Fig.~\ref{fig:dynamics}             & baseline (no changes)\\
Fig.~\ref{fig:phase}A               & overlay: $s_H = 0.6$ vs.\ $0.9$ \\
Fig.~\ref{fig:phase}B               & $s_H \in [s_L, 1]$ swept, $I_H = I_L$ at baseline \\
Fig.~\ref{fig:phase}C               & $s_H \in [s_L, 1]$, $I_H = I_L \in [0.5,1]$ swept\\
Fig.~\ref{fig:phase}D               & overlay: $\rho = 0.2$ vs.\ $0$ \\
Fig.~\ref{fig:phase}E               & as Fig.~\ref{fig:phase}B with $\rho = 0.2$\\
Fig.~\ref{fig:phase}F               & as Fig.~\ref{fig:phase}C with $\rho = 0.2$\\
Fig.~\ref{fig:delay}                 & $\mathbf{C}_0 = (0,0,0.5,0.5)$, $s_H \in \{0.6, 0.7, 0.9\}$, $I_H(t) = \min(\max((t-20)/60, 0), 1)$, $I_L = 0.5$, $t_{\text{final}} = 100$ \\
Fig.~\ref{fig:dynamic_connectivity} & $\rho(t) = 0$ for $t < 50$; $\rho \in \{0.2, 0.5\}$ for $t \geq 50$; $t_{\text{final}} = 100$ \\
Fig.~\ref{fig:s_drop}               & $\mathbf{C}_0 = (0,0,0.5,0.5)$, $s_H(t) = 0.9$ for $t < 50$ and $s_H \in \{0.8, 0.7, 0.6\}$ for $t \geq 50$; $I_H = 1.0$, $I_L = 0.5$, $t_{\text{final}} = 100$ \\
\midrule
Fig.~\ref{fig:phase_SI}A            & as Fig.~\ref{fig:phase}C, swept $I \in [0, 1]$ \\
Fig.~\ref{fig:phase_SI}B            & as Fig.~\ref{fig:phase}F, swept $I \in [0, 1]$ \\
Fig.~\ref{fig:delay_SI}      & as Fig.~\ref{fig:delay}, $I_H(t)=1$ \\
Fig.~\ref{fig:connectivity_SI}      & as Fig.~\ref{fig:dynamic_connectivity} \\
\bottomrule
\end{tabular}
\end{table}

\section{Further model analysis}

\subsection{Equilibrium priority across the full range of objective severity}

The analysis so far has considered objectively severe issues, those for which a majority of individuals would be concerned under purely individual learning. Here we extend the analysis to issues with severity below 0.5, meaning that fewer than half of individuals would be concerned when learning individually. Misprioritization persists in this regime, but its direction reverses. Once severity falls below 0.5, the more socially learned issue becomes overprioritized.

\begin{figure}[H]
    \centering
    \includegraphics[width=\linewidth]{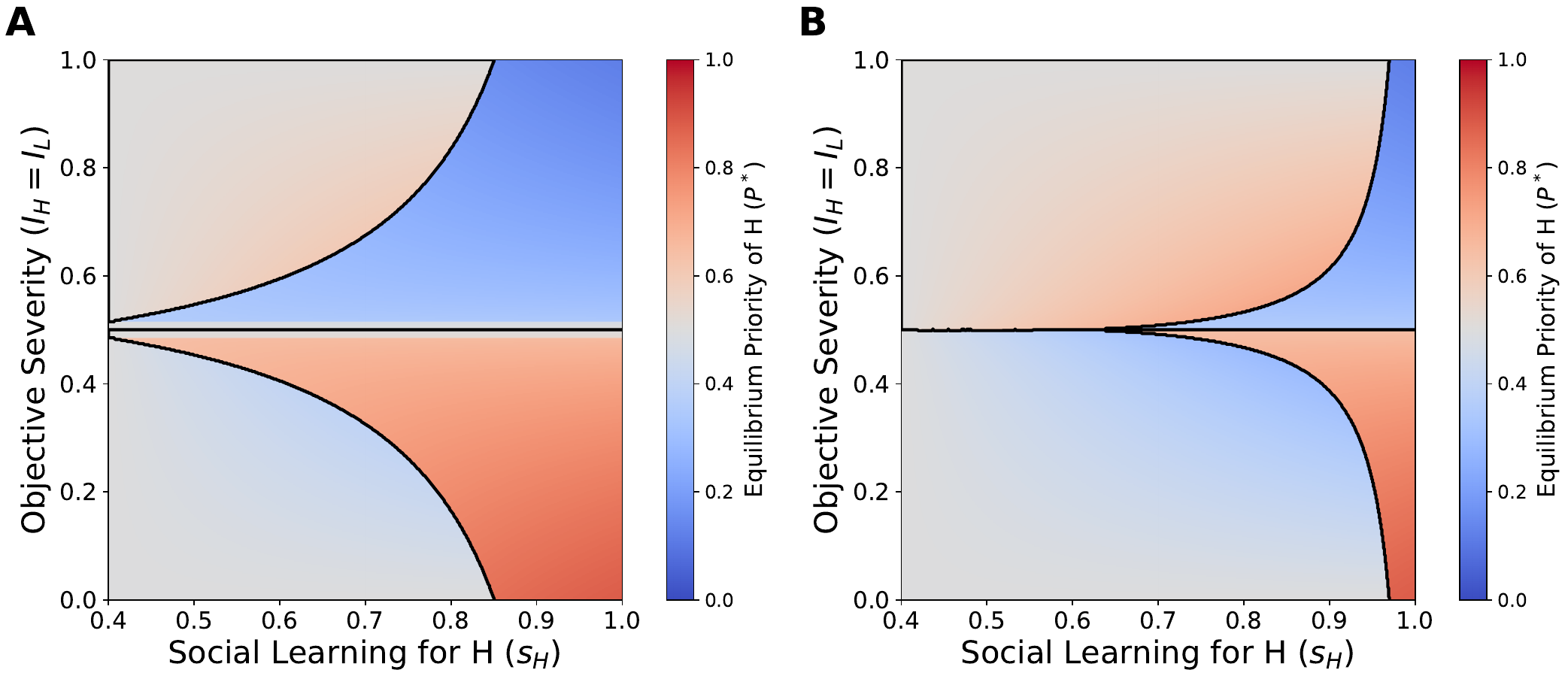}
    \caption{Equilibrium priority of $H$ across the full range of objective severity, extending Fig.~\ref{fig:phase}C and F. (A) Without intergroup learning ($\rho = 0$). (B) With intergroup learning ($\rho = 0.2$). Upper halves of the panels appear in the main text. Lower halves show that at low severity ($I_H = I_L < 0.5$), $H$ is overprioritized through social reinforcement, the mirror image of the high-severity regime.}
    \label{fig:phase_SI}
\end{figure}

Figure~\ref{fig:phase_SI} extends the equilibrium maps of Fig.~\ref{fig:phase}C and F to the full severity range $I_H = I_L \in [0,1]$. The upper halves of the panels reproduce the main-text maps, where $H$ is underprioritized at high social learning. At low severity ($I_H = I_L < 0.5$), the pattern reverses, and $H$ is overprioritized through social reinforcement when only a minority would be concerned under purely individual learning. Intergroup learning shrinks the region of misprioritization (both under- and overprioritization).

\subsection{Evolution of priority with severity held constant at its maximum}

\begin{figure}[H]
  \centering
  \includegraphics[width=.8\linewidth]{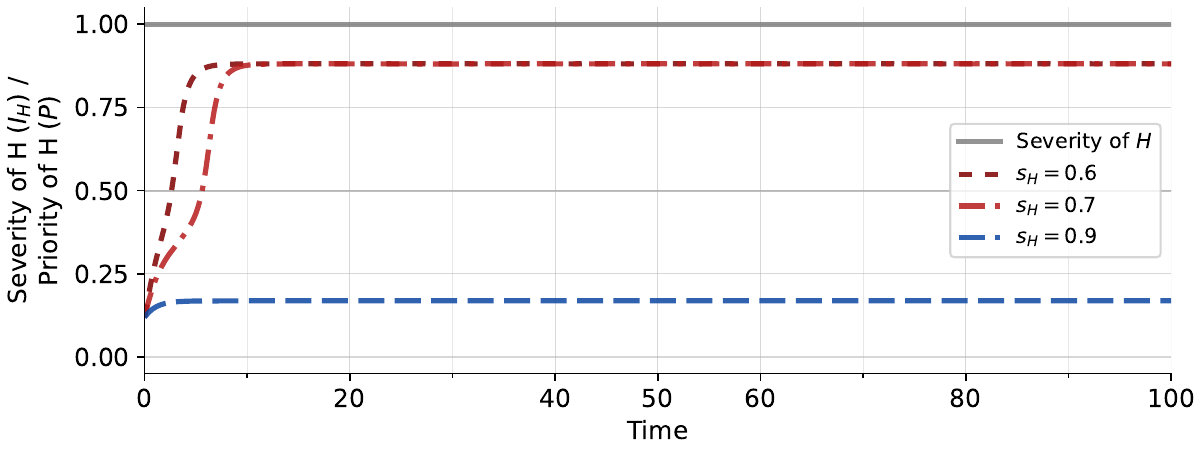}
  \caption{With severity held constant at its maximum, priority reaches equilibrium within 10 time units. The gray line shows the objective severity of $H$ fixed at $1$, while $L$ is held at $0.5$. Colored lines show the priority of $H$ over $L$ at three levels of social learning.}
\label{fig:delay_SI}
\end{figure}

A gradual increase in the objective severity of an issue delays reprioritization, in contrast to the case in which severity begins at a high level and is sustained there. Figure~\ref{fig:delay} illustrates this delay when the objective severity of issue $H$ increases gradually while that of issue $L$ is fixed. Figure~\ref{fig:delay_SI} shows the corresponding result with the objective severity of issue $H$ fixed at 1 from the outset, and reprioritization occurs with far less delay (within 10 time units). Comparing the two figures shows that whether severity starts at its maximum or rises gradually to it affects only the lag in reaching the equilibrium priority, not the equilibrium level itself, which is determined by social learning in this analysis.

\subsection{Group-level dynamics under increases in intergroup learning}

\begin{figure}[H]
    \centering
    \includegraphics[width=0.7\linewidth]{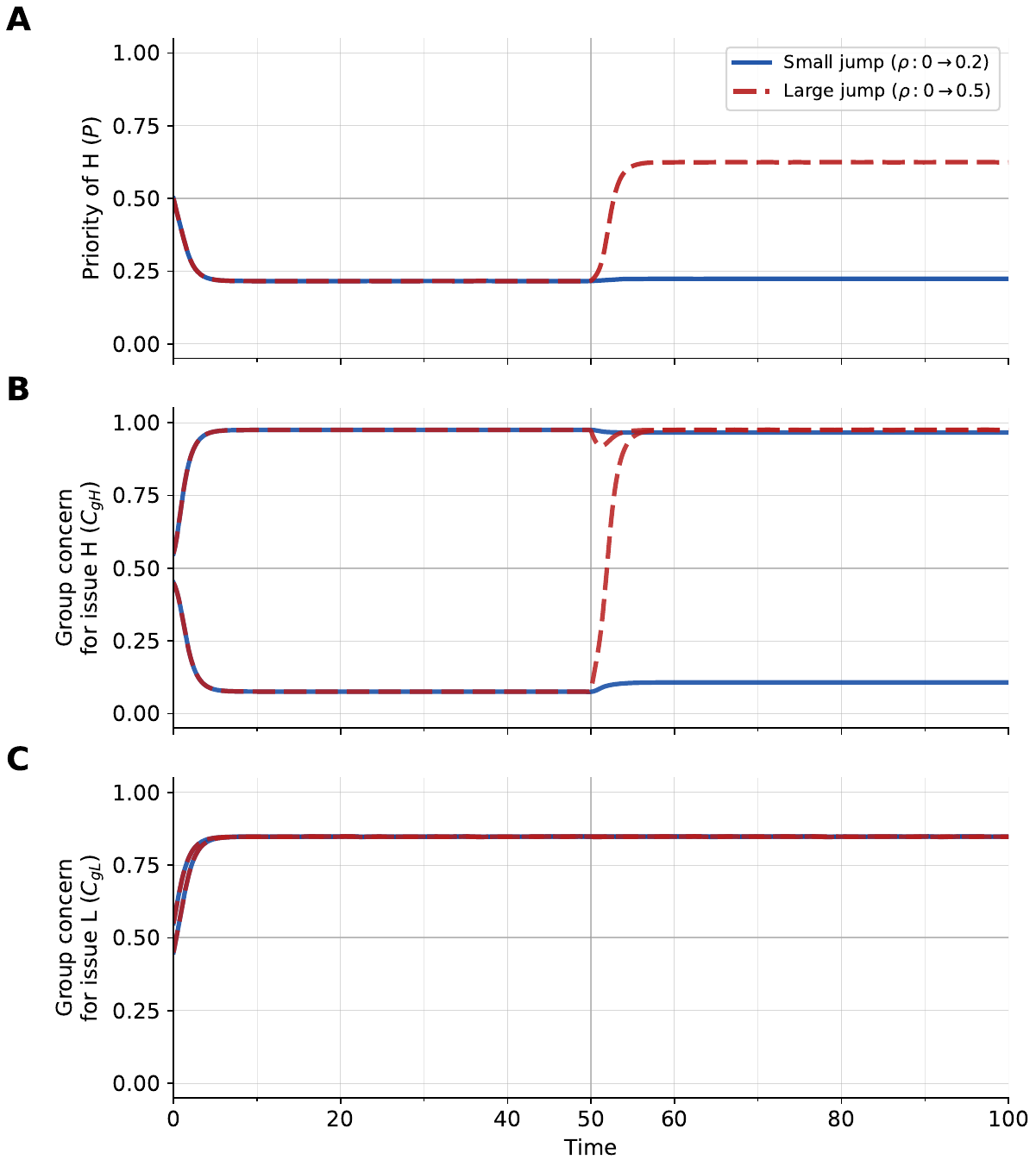}
    \caption{Group-level dynamics underlying Fig.~\ref{fig:dynamic_connectivity}. At $t = 50$, intergroup learning increases from $\rho = 0$ to either $0.2$ (small jump) or $0.5$ (large jump). (A) Population priority for $H$ over $L$. (B) Group-level concern for issue $H$ in each group, showing that only the large jump collapses between-group divergence. (C) Group-level concern for issue $L$, which remains uniformly high in both scenarios.}
    \label{fig:connectivity_SI}
\end{figure}

Figure~\ref{fig:connectivity_SI} displays the group-level concern dynamics underlying the threshold behavior in Fig.~\ref{fig:dynamic_connectivity}. Before the perturbation at $t = 50$, concern for $H$ is polarized between groups while concern for $L$ is uniformly high. A large enough increase in intergroup learning drives the polarized concern for $H$ to converge toward a high level in both groups, whereas a small jump only slightly reduces the magnitude of the divergence. The small jump therefore leaves the polarization in concern for $H$ largely intact, while the large jump collapses it, producing the priority shift seen in Fig.~\ref{fig:dynamic_connectivity}A. Concern for $L$ is unaffected by either intervention.

\printbibliography